\documentclass[journal=jacsat,manuscript=article]{achemso}
\setkeys{acs}{super = true}
\setcitestyle{super,open={},close={}}

\usepackage{chemformula} 
\usepackage[T1]{fontenc} 
\usepackage[nomarkers,figuresonly]{endfloat}

\author{Giuseppe Antonacci}
\altaffiliation{The authors contributed equally to the work}
\affiliation[UGent]
{Photonics Research Group, INTEC, Ghent University-imec, Ghent 9052, Belgium}
\alsoaffiliation[NB]
{Center for Nano- and Biopothonics (NB-Photonics), Ghent University, 9052 Ghent, Belgium}
\email{giuseppe.antonacci@ugent.be}

\author{Jeroen Goyvaerts}
\altaffiliation{The authors contributed equally to the work}
\affiliation[UGent]
{Photonics Research Group, INTEC, Ghent University-imec, Ghent 9052, Belgium}
\alsoaffiliation[NB]
{Center for Nano- and Biopothonics (NB-Photonics), Ghent University, 9052 Ghent, Belgium}

\author{Haolan Zhao}
\affiliation[UGent]
{Photonics Research Group, INTEC, Ghent University-imec, Ghent 9052, Belgium}
\alsoaffiliation[NB]
{Center for Nano- and Biopothonics (NB-Photonics), Ghent University, 9052 Ghent, Belgium}

\author{Bettina Baumgartner}
\affiliation[Vienna]
{Institute of Chemical Technologies and Analytics, TU Wien, A-1060 Vienna, Austria}

\author{Bernhard Lendl}
\affiliation[Vienna]
{Institute of Chemical Technologies and Analytics, TU Wien, A-1060 Vienna, Austria}

\author{Roel Baets}
\affiliation[UGent]
{Photonics Research Group, INTEC, Ghent University-imec, Ghent 9052, Belgium}
\alsoaffiliation[NB]
{Center for Nano- and Biopothonics (NB-Photonics), Ghent University, 9052 Ghent, Belgium}

\title[Gas sensor]
  {Ultra-sensitive refractive index gas sensor with functionalized silicon nitride photonic circuits}


\begin{document}

\makeatletter
\setlength\acs@tocentry@height{10.5cm}
\setlength\acs@tocentry@width{5cm}
\makeatother

\begin{tocentry}
\centering
 \includegraphics{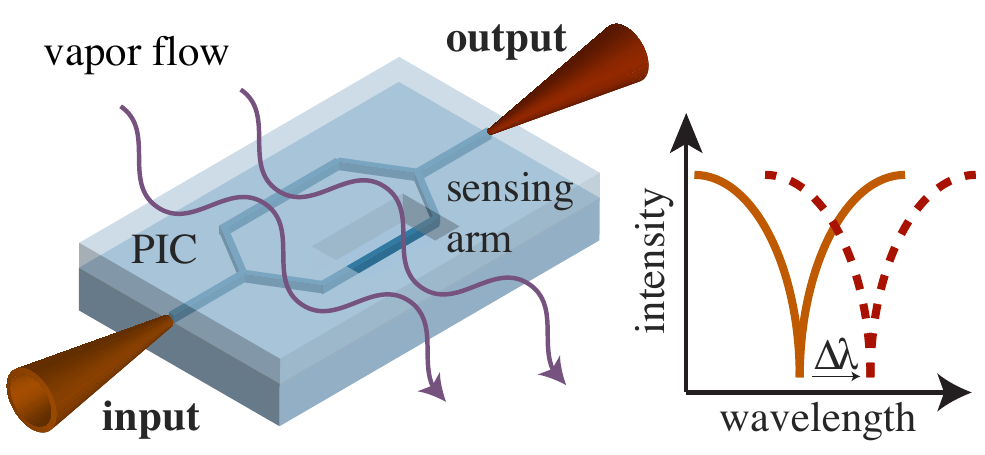}

\end{tocentry}

\begin{abstract}

Portable and cost-effective gas sensors are gaining demand for a number of environmental, biomedical and industrial applications, yet current devices are confined into specialized labs and cannot be extended to general use. Here, we demonstrate a part-per-billion-sensitive refractive index gas sensor on a photonic chip based on silicon nitride waveguides functionalized with a mesoporous silica top-cladding layer. Low-concentration chemical vapors are detected by monitoring the output spectral pattern of an integrated Mach-Zehnder interferometer having one coated arm exposed to the gas vapors. We retrieved a limit of detection of 65$\,$ppb, 247$\,$ppb and 1.6$\,$ppb for acetone, isopropyl alcohol and ethanol, respectively. Our on-chip refractive index sensor provides, to the best of our knowledge, an unprecedented sensitivity for low gas concentrations based on photonic integrated circuits. As such, our results herald the implementation of compact, portable and inexpensive devices for on-site and real-time environmental monitoring and medical diagnostics.


\end{abstract}

Gas sensing is a central task for a plethora of applications including medical diagnostics, pollution monitoring and industrial quality process control. In medical diagnostics, for example, recent studies have suggested the use of gas sensors in breath analysis to detect diseases such as lung cancer\cite{di2003lung} and diabetes\cite{ryabtsev1999application}. Environmental safety monitoring has also been a topic of recent interest in the microelectronics sensing industry with regards to the miniaturization and cost benefits that go along with replacing the current technologies. Despite the presence of broad technological solutions for gas detection\cite{ho2001review, jimenez2007gas, ritari2004gas}, these typically rely on bulk devices that lack the affordability, robustness and selectivity requirements for ensuring a global implementation outside specialized labs. 

Demand for compact, cost-effective, purely passive and portable gas sensors has instigated new research on chip-scale silicon photonic devices. These promise benefits in terms of cost, size and energy efficiency over traditional technologies that would potentially enable a wide spread and on-site adoption. In this regard, the novel silicon nitride (SiN) platform offers an excellent solution for sensing as a consequence of its broadband transparency window into the visible domain and the generally lower propagation losses compared to the SOI platform \cite{rahim2017expanding}. Given its versatility and the possibility to integrate multiple optical functionalities, the SiN platform has already enabled the implementation of several sensing devices for a variety of applications including  biosensing \cite{subramanian2015silicon, martens2018low, claes2009label}, Raman spectroscopy \cite{dhakal2014evanescent, raza2019high, zhao2018stimulated} and absorption spectroscopy \cite{ryckeboer2014glucose}. More recently, the SiN platform has further seen adoption for gas sensing. Celo et. al. \cite{celo2009interferometric} first demonstrated the use of an integrated Mach-Zehnder interferometer (MZI) with SiN rib-waveguide structures coated with $\sim1\mu$m ZnO and TiO2 films to probe N2 and CO2. Similar approaches based on integrated MZIs realized with SOI and SiN platforms have been demonstrated with different types of coatings and waveguide structures to detect organic solvents \cite{fabricius1992gas}, ethanol vapors \cite{ghosh2017compact} and methane \cite{dullo2015sensitive} reporting a limit of detection (LOD) in the order of  tens of ppm. On-chip refractive index gas sensing has been further demonstrated through microring resonators \cite{robinson2008chip, passaro2007ammonia, orghici2010microring, yebo2010integrated}, where a LOD down to 5 ppm has been recently achieved with acidic nano-porous aluminosilicate films for ammonia sensing \cite{yebo2012selective}. Other emerging integrated approaches include absorption spectroscopy \cite{stievater2014trace,tombez2017methane} and Raman spectroscopy \cite{holmstrom2016trace, tyndall2018waveguide}. These methods promise high selectivity, yet the requirement of signal-enhancement polymer coatings still limits their detection to specific binding analytes and to long enrichment times due to bulk diffusion. 
 
 In this Letter, we demonstrate a part-per-billion-sensitive on-chip gas sensor on a low-loss silicon nitride platform. Our sensing mechanism hinges on the refractive index change of the SiN waveguides evanescently exposed to the gas analytes across the openclad arm of an integrated MZI. To enhance the instrumental sensitivity, the sensing arm of the MZI was coated with a functionalized mesoporous silica layer that preferentially adsorbs organic solvents. The light traveling along the exposed waveguides experiences a different refractive index when a gas vapor is adsorbed by the top-cladding mesoporous layer, resulting in a wavelength shift of the interference fringes generated by the MZI. We demonstrate this concept for a number of gas vapors including acetone, isopropyl alcohol (IPA) and ethanol. 

 
 \begin{figure}[htbp]
 \centering
 \includegraphics[width=4in]{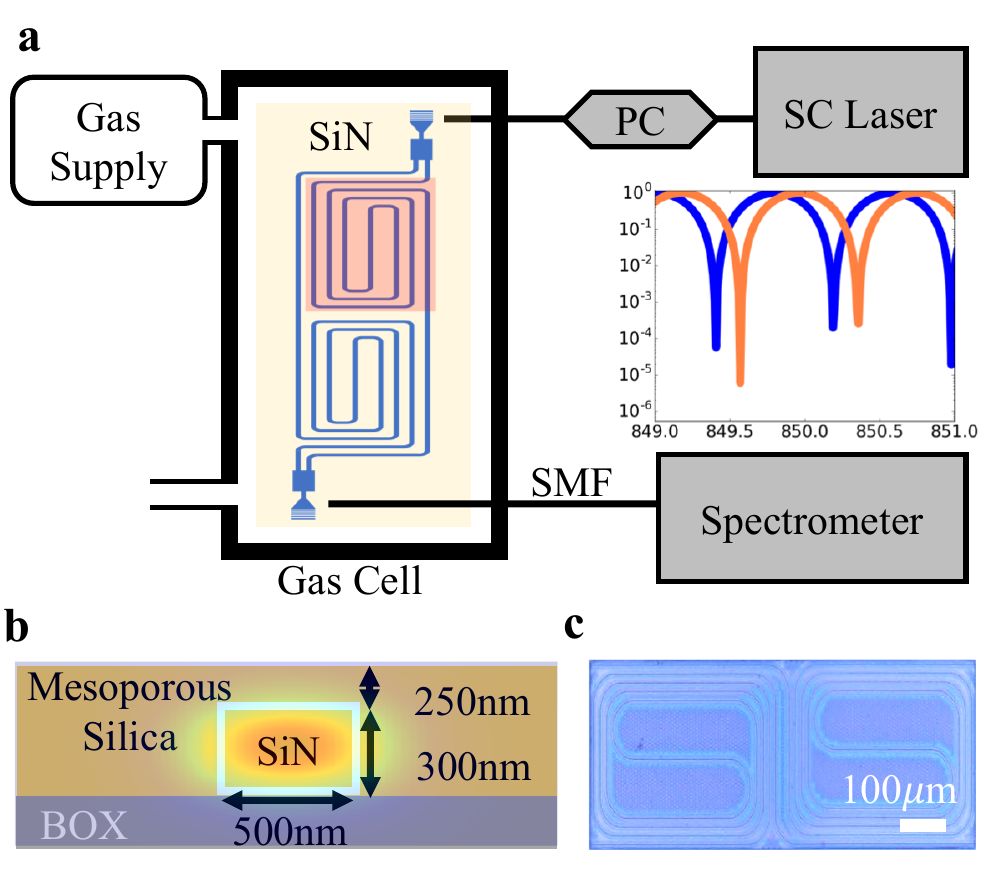}
  \caption{Experimental setup (a). A supercontinuum (SC) laser beam was coupled into the SiN waveguides to probe the injected gas vapors. A polarization controller (PC) was used to ensure a TE mode propagating along the waveguides. The spectral readout was performed in real-time by a spectrometer. Cross section of our 300$\,$nm SiN waveguides with functionalized mesoporous coating (b) and image of the integrated MZI (c).}
  \label{fig1}
\end{figure}

Figure \ref{fig1}a illustrates the experimental setup. A supercontinuum (SC) laser source (NKT-EXR-4) was used in all experiments. The laser beam was directly coupled from a single mode fiber  into the SiN waveguides of our photonic integrated circuit (PIC) through a grating coupler, which was designed for a $850\,$nm central wavelength to further enable the integration of high power VCSELs.  The PIC was fabricated on imec's BioPIX  platform (Figure \ref{fig1}b), which provides a waveguide core thickness of 300 nm and is suited for applications targeting the near-infrared (700-900$\,$nm) range. The SiN used here is a PECVD nitride core of 1.89 refractive index, providing a relatively low material contrast with the 1.46 index of the SiO$_2$ cladding. This low material index contrast is particularly suited for sensing, as the mode results to be less confined across the waveguide core resulting in an extended evanescent tail, and therefore in a longer interaction length with the adsorbed gas analytes. As further demonstrated through the EU PIX4Life pilot line \cite{Pix4Life}, the BioPIX SiN platform is also well-suited to design complex photonic components, including dual-etch staircase grating couplers, splitter modules and low-loss on-chip spectrometers. 

The main building-block of the sensor is the integrated MZI (Figure \ref{fig1}c). Given the low-loss (typically $\sim0.5\,$dB/cm) of our SiN waveguides, we were able to design the MZI with long interference arms of $\sim5\,$mm length to increase the overall optical path length. This, in turn, enhanced the device sensivity without detrimental influence on the signal contrast. The Free Spectral Range (FSR) of the MZI was designed to be around 1 nm for low-index coatings, providing the opportunity for further electro-optic integration of narrowband tunable lasers. Moreover, the strip waveguides of the sensing arm of the MZI were functionalized with an ordered mesoporous silica coating layer of $\sim500\,$nm thickness and with a 3D hexagonal structure \cite{baumgartner2018situ, baumgartner2019pore}. Assuming a porosity of $\sim50\%$ and using the Bruggeman effective medium approximation \cite{doi:10.1002/andp.19354160705}, we estimated an effective medium refractive index of 1.21 in air. Moreover, the film was functionalized with hexamethyldisilazane to increase the hydrophobicity of the film \cite{baumgartner2020mesoporous}. This allowed for excluding the strong absorber water, while enhance sensitivity towards organic molecules. 

The use of a SC laser source enabled the employment of a broadband spectral analyzer (Agilent OSA 86140B) for direct and real-time readout of the output spectral interference pattern with a typical exposure time of 500$\,$ms. The polarization of the SC source was controlled using a linear polarizer and a polarization controller acting as a half-wave plate to ensure coupling of the TE mode into the waveguides and to maximize the optical signal strength. Moreover, the PIC was kept enclosed inside a custom gas cell having an input and an output port for gas vapor flowing, and a transparent window to enable the optical incoupling and outcoupling. A gas generator (VICIMetronics Model 505) was employed to inject low (0.1$\,$ppb-100$\,$ppm) gas concentrations into the gas cell through a controlled permeation mechanism (Supporting Figure 1a).


 \begin{figure}[htbp]
 \centering
 \includegraphics[width=4in]{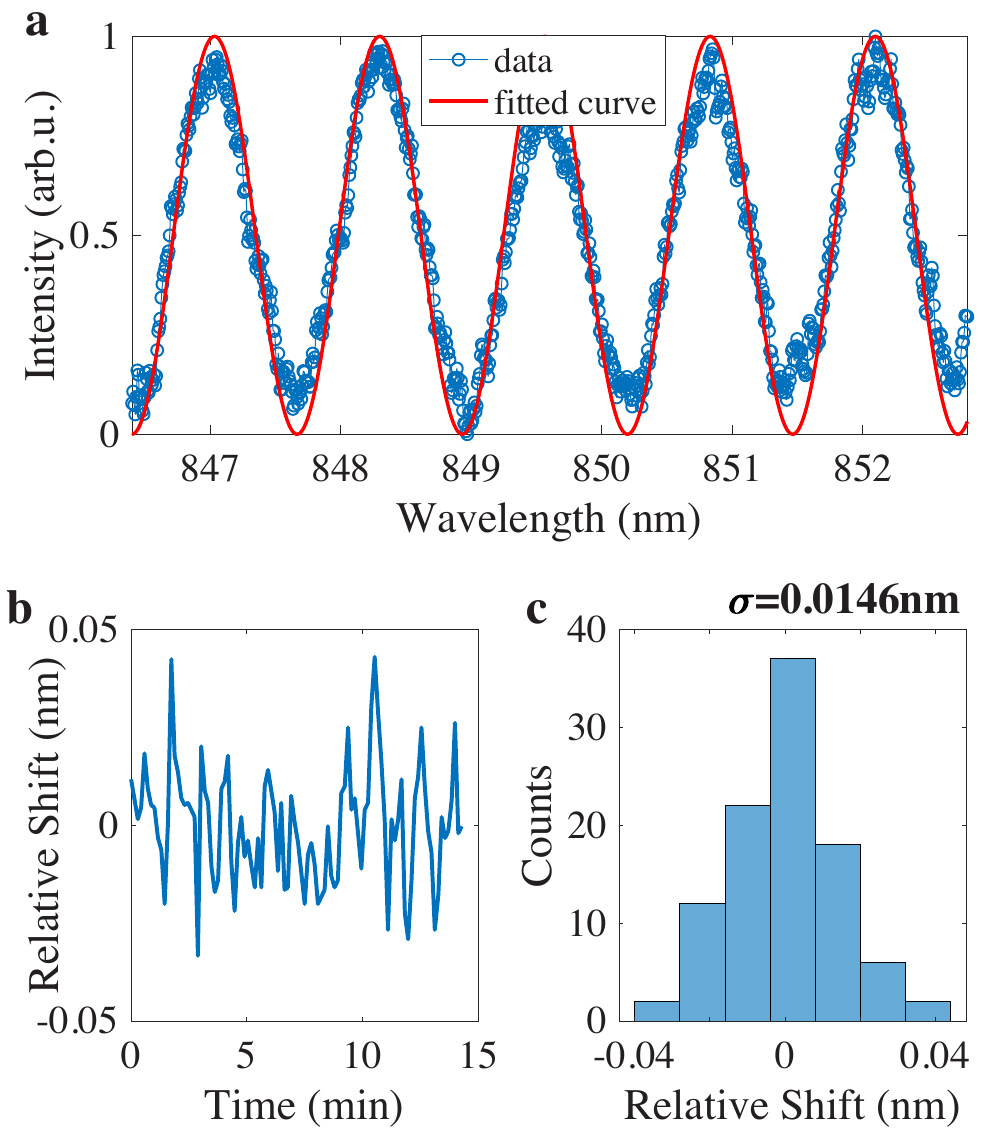}
  \caption{Spectral interference pattern of the integrated MZI fitted with a sinusoidal function (a). The wavelength variation of a fringe was monitored over a time period of 15 mins to assess the sensor stability (b). Histogram of the data set, which gave a sigma of $\sigma = 0.0146 \,$nm as a baseline noise level on the extracted wavelength (c).}
  \label{fig2}
\end{figure}

The sensor was first optically characterized to measure the device sensitivity and the optical stability in controlled environmental conditions. In more detail, the output spectral interference patterns of the MZI were acquired over time by the spectrometer while the photonic chip was kept inside the gas cell at room temperature ($22^{\circ}$C) with no gas vapor flowing. The output intensity profiles were fitted using a suitable sinusoidal function to track the location of the interference fringes of the MZI, as illustrated in Figure \ref{fig2}a.  Figure \ref{fig2}b and c show a plot of the location in the spectral domain of an arbitrary interference fringe as a function of time and a histogram of the same dataset for a total of $N\sim100$ points, respectively. Results show good stability of the sensor over a time period of $\sim$15 mins, which was comparable to the data acquisition time period in our sensing experiments. From these measurements, we estimated a sigma of $\sigma = 0.0146 \,$nm, which sets the baseline noise level of our gas sensor. 

 \begin{figure}[htbp]
 \centering
 \includegraphics[width=4in]{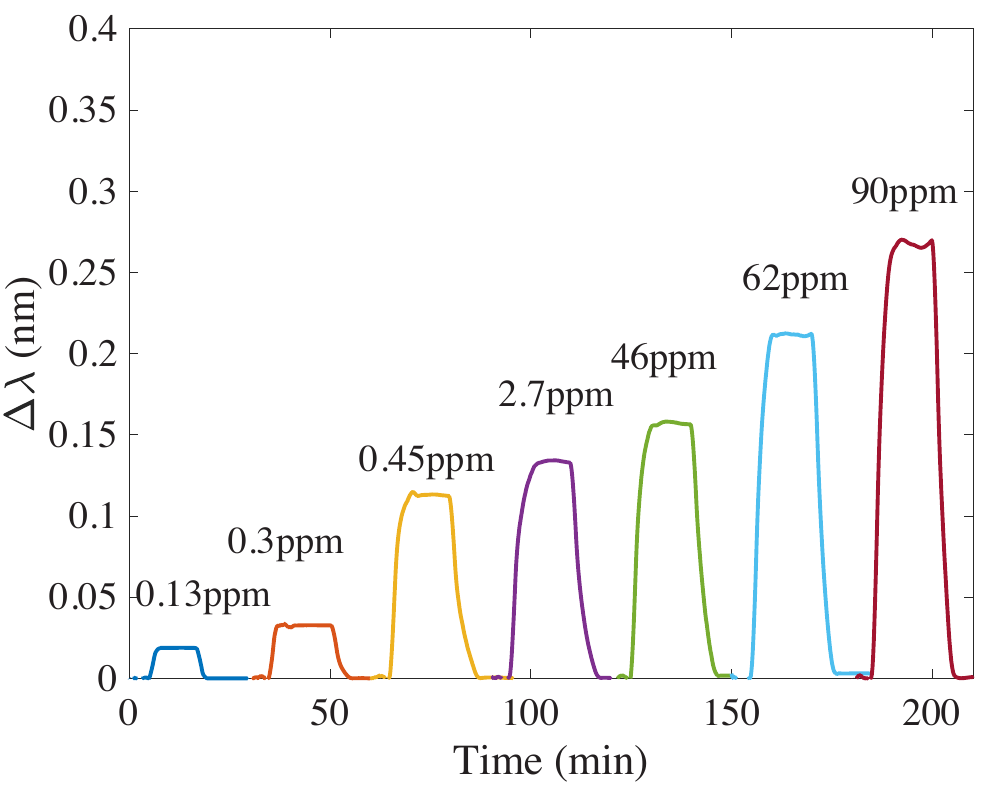}
  \caption{Representative response curves of our gas sensor to the exposure to acetone vapors at different concentrations. Gas detection was accomplished by tracking the wavelength variation associated to a reference MZI fringe over time. Results show a clear response even at concentrations below a part-per-million scale.}
  \label{fig3}
\end{figure}

Figure \ref{fig3} shows the sensor response in terms of the relative wavelength shift over time after the exposure to acetone  at different concentrations. To supply a broad range of concentrations to the gas cell, the gas generator was tuned either in temperature (setting the amount of gas permeation inside the supplier chamber) or in the diluent flow. For each measurement, the gas sensor was first exposed to the low gas concentration flow for 15 minutes to ensure a full adsorption cycle and thus a stabilization of the output spectral interference pattern. This process was then followed by an air diluent reflow for an equal amount of time to return the mesoporous coating to its original state upon desorption and the optical signal to its initial baseline. The spectral interference patterns at the output of the MZI were recorded continuously by the spectrometer with time steps of 1 sec and fitted to finely track the wavelength shift of an arbitrary reference fringe. Results show clear visible signal responses of our gas sensor even at low (<1$\,$ppm) gas concentrations, with spectral shifts that were well above the baseline noise level of the device.

Generally speaking, MZIs are extremely sensitive devices to thermodynamic and environmental changes, including variations in temperature, pressure and humidity. 
To ensure that the observed spectral shifts were given by the adsorption of the gas analytes rather than by a change in temperature caused by the permeation mechanism of the gas generator, we measured the sensor response at a fixed chamber temperature to the exposure of both normal air flow and acetone. Results illustrated in Supporting Figure 1b gave convincing evidence of the negligible influence of the gas sensor on the heating mechanism occurring inside the permeation chamber of the gas generator.  
A similar test was performed to investigate the effect of humidity on the gas sensor (Supporting Figure 2). As we expected, a change in the environmental humidity was reflected in a small step variation of the MZI fringes. However, exposure of gas vapors in altered humidity levels had no significant change in the relative spectral shift as compared to that given with the original humidity level. 

 \begin{figure}[htb]
 \centering
 \includegraphics[width=4in]{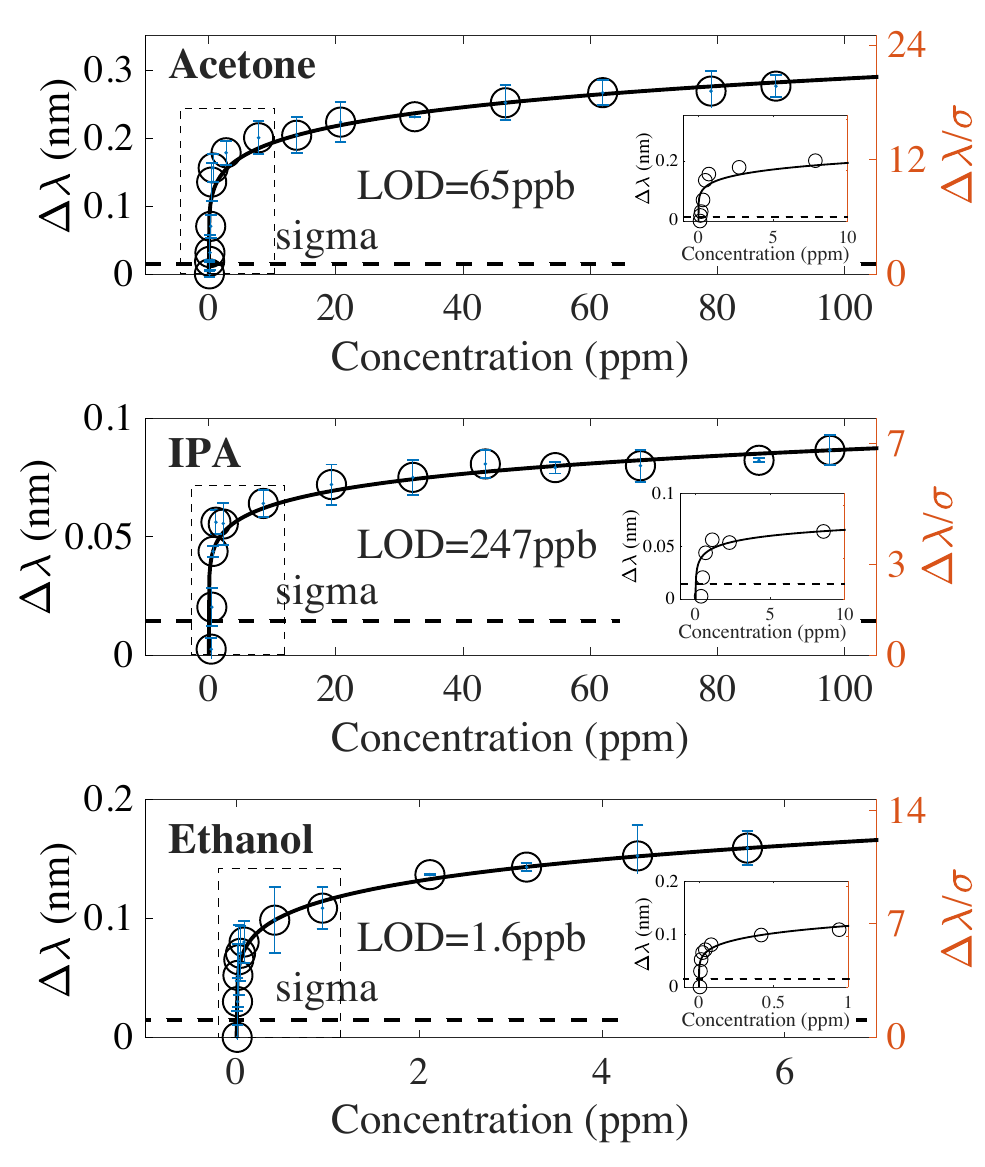}
  \caption{Relative spectral variations of the MZI interference fringes as a function of gas concentrations. Data fitting was performed using the Freundlich model for multilayer mesoporous coatings. The LOD for acetone, IPA and ethanol vapors was estimated to be 65$\,$ppb, 247$\,$ppb and 1.6$\,$ppb, respectively. Error bars come from the std of wavelength shift values obtained upon completion of the adsorption cycle. (Inset) Low-concentration plots.}
  \label{fig4}
\end{figure}

To investigate the LOD of our gas sensor, we swept the concentration of a number of gas vapors namely acetone, IPA and ethanol, and record the wavelength shift of the MZI fringes. Figure \ref{fig4} resumes the response curves of the gas sensor for the tested gas vapors. Wavelength shift values were obtained by averaging the spectral shifts given upon completion of the adsorption process. As shown, the sensor had a non-linear response to the exposure of gas concentrations. This behavior can be described by the Freundlich model, which involves a multilayer adsorption with decreasing adsorption enthalpy in mesoporous silica \cite{baumgartner2018situ}. Using this model to fit the data and to project the obtained curves to the measured instrumental sensitivity defined by the sigma, we obtained a LOD of acetone, IPA and ethanol to be 65$\,$ppb, 247$\,$ppb and 1.6$\,$ppb respectively. The difference in the measured LOD might be due to a number of reasons, such as the different vapor pressures and refractive index of the injected gas vapors as well as to the amount of gas molecules adsorbed by the mesoporous film coating. 

In conclusion, we demonstrated an integrated ppb-sensitive refractive index gas sensor with functionalized SiN photonic circuits. Our results show, to the best of our knowledge, the lowest reported LOD with integrated gas sensor devices on a silicon nitride platform. The LOD may be further improved by enhancing the stability of the sensor against small mechanical drifts or thermodynamic variations so that the instrumental sensitivity would only be limited by the  noise and ultimately by the spectral resolution of the detector. Despite the unprecedented LOD in a SiN chip, our integrated MZI sensor is still limited by an intrinsically low selectivity that does not allow to gather information about the detected gas analyte. As such, further research will focus on advanced mesoporous coatings that can selectively adsorb  specific gas analytes depending on their engineered porous size and functionalization. This may further enable multiplexed assays exploiting multiple MZIs on a single chip. In parallel, progress on signal enhancement on chip-scale Raman spectroscopy \cite{zhao2020multiplex} may also provide an effective and complimentary approach for gas sensing. 
The maturity and flexibility of the SiN platform will allow for further electro-optic integration in the future by combining light sources and detectors on chip, paving the way to fully functionalized, compact and portable gas sensing devices enabling new exciting ventures for environmental safety, healthcare and industrial processing. 

\begin{acknowledgement}

The authors thank the EU pilot lines PIX4Life and PIXAPP for continuous support and funding. We further acknowledge funding from AQUARIUS project, which has received funding from the EU Horizon 2020 research and innovation program under grant agreement No. 731465.

\end{acknowledgement}

\bibliography{ACS.bib}




\end{document}